\documentclass[pre,twocolumn,superscriptaddress,showpacs]{revtex4-1}
\usepackage[breaklinks]{hyperref}
\usepackage[hyphenbreaks]{breakurl}
\usepackage{amsfonts, amssymb, amsmath}
\usepackage{psfrag}
\usepackage{graphicx}
\usepackage{color}
\usepackage{soul}

\usepackage{todonotes}

\makeatletter
\newcommand*{\rom}[1]{\expandafter\@slowromancap\romannumeral #1@}
\makeatother

\makeatletter
\newcommand*\bigcdot{\mathpalette\bigcdot@{.5}}
\newcommand*\bigcdot@[2]{\mathbin{\vcenter{\hbox{\scalebox{#2}{$\m@th#1\bullet$}}}}}
\makeatother

\newcommand*{\E}{\mathrm{e}}

\begin{document}
\title{Microscopic Cross-Correlations in the Finite-Size Kuramoto Model of Coupled Oscillators}
\date{\today}
\author{Franziska Peter}
\affiliation{Institute of Physics and Astronomy, University of Potsdam, Karl-Liebknecht-Stra\ss{}e 24-25, 14476 Potsdam, Germany}
\author{Chen Chris Gong}
\affiliation{Institute of Physics and Astronomy, University of Potsdam, Karl-Liebknecht-Stra\ss{}e 24-25, 14476 Potsdam, Germany}
\author{Arkady Pikovsky}
\affiliation{Institute of Physics and Astronomy, University of Potsdam, Karl-Liebknecht-Stra\ss{}e 24-25, 14476 Potsdam, Germany}
\affiliation{Department of Control Theory, Nizhny Novgorod State University, Gagarin Av.\ 23, 606950, Nizhny Novgorod, Russia}

\begin{abstract}
Super-critical Kuramoto oscillators with distributed frequencies separate into two disjoint groups: an ordered one locked to the mean field, and a disordered one consisting of effectively decoupled oscillators -- at least so in the thermodynamic limit. In finite ensembles, in contrast, such clear separation fails: The mean field fluctuates due to finite-size effects and thereby induces order in the disordered group. To our best knowledge, this publication is the first to reveal such an effect, similar to noise-induced synchronization, in a purely deterministic system. We start by modeling the situation as a stationary mean field with additional white noise acting on a pair of unlocked Kuramoto oscillators. An analytical expression shows that the cross-correlation between the two increases with decreasing ratio of natural frequency difference and noise intensity. In a deterministic finite Kuramoto model, the strength of the mean field fluctuations is inextricably linked to the typical natural frequency difference. Therefore, we let a fluctuating mean field, generated by a finite ensemble of active oscillators, act on pairs of passive oscillators with a microscopic natural frequency difference between which we then measure the cross-correlation, at both super- and sub-critical coupling.

\end{abstract}

\maketitle
\section{Introduction}
Synchronization -- the mutual adjustment of frequencies among weakly coupled 
self-sustained oscillators --
is a prominent example of the emergence of order in out-of-equilibrium 
systems in physics, engineering, biology, and
other fields~\cite{Kuramoto1984,Strogatz-03,pikovsky2003synchronization}. 
In large ensembles, it appears as a non-equilibrium phase transition, 
where the organizing
action of sufficiently strong mutual coupling wins over the disorganizing action of to the 
diversity in natural frequencies. 
 The paradigmatic
model of this phenomenon, created by Kuramoto~\cite{Kuramoto1984,Acebron-etal-05}, is fully solvable in the 
thermodynamic limit~\cite{Kuramoto1984,Ott2008}. The characteristic feature of the Kuramoto-type
synchronization transition is the coexistence of two subgroups of oscillators in the partially synchronized state: 
the oscillators in the ordered group are locked by the mean field
and coherently contribute to it, while the disordered units are not 
locked and rotate incoherently. 
With increasing coupling strength, the former group grows in size, 
as more and more oscillators are
locked by the increasing mean field.

The qualitative features, established in the thermodynamic limit, remain approximately valid for
finite ensembles. Here, similar to finite-size effects in equilibrium phase 
transitions, the order parameter, i.~e.~the macroscopic mean field, fluctuates with an amplitude that depends on the ensemble size in a 
nontrivial 
way~\cite{Daido1987, Hong2007, Nishikawa2014,Hong_etal-15,FPAP2018}.
 These fluctuations are most pronounced close to the criticality, and can 
be attributed to weak chaoticity
of the finite population dynamics ~\cite{LE, Gilad2013}.

The goal of this paper
is to show that the finite-size fluctuations of the mean field
have an additional effect on the population -- quite counter-intuitively an ordering effect:
the disordered oscillators become correlated pairwise, while in the thermodynamic limit the cross-correlations
disappear. Below, we consider two basic setups to show this phenomenon. 
First, we study a population
in the thermodynamic limit, but with the mean field being subject to external
white noise fluctuations (Section \ref{sec:setup1}). 
This ideal setup allows for an analytic solution, showing
the dependence of the cross-correlation between the oscillators
on the fluctuation intensity and on 
the natural frequency difference.
In the second setup (Section \ref{sec:setup2}), we numerically quantify the cross-correlation due to the intrinsic 
finite-size-induced fluctuations of the mean field, first for super- than for sub-critical coupling. 
This latter case is similar
to other organizing macroscopic manifestations of finite-size fluctuations such as
finite-size-induced phase transitions~\cite{Pikovsky-Rateitschak-Kurths-94,Komarov-Pikovsky-15b} and stochastic 
resonance~\cite{Pikovsky-Zaikin-Casa-02}. In fact, this ordering action of 
finite-size fluctuations can be qualitatively traced to the effect of synchronization
by common noise, known for identical and nonidentical oscillators, which are either 
coupled or uncoupled~\cite{Goldobin-Pikovsky-04,Goldobin-Pikovsky-05b,Nagai_Kori_2010,Goldobin2016}.

\section{Mean field with external fluctuations in thermodynamic limit} \label{sec:setup1}
\subsection{Stationary mean field}
Before discussing the mean-field model with external fluctuations, 
we first quantify ordered and disordered states in the Kuramoto
model of mean-field coupled oscillators in the thermodynamic limit where no fluctuations are present. 
The model is formulated as follows: Oscillators
are described by their phases $\varphi$, and are coupled via the complex 
mean field $Z\equiv R\E^{i\Phi} = \int_{0}^{2\pi} d\varphi \int_{-\infty}^{\infty} d\Omega
\,P(\varphi| \Omega) g(\Omega) \E^{i\varphi}  $ as
\begin{equation}
\dot\varphi=\Omega+\varepsilon \text{Im}(Z\E^{-i\varphi})\;.
\label{eq:km}
\end{equation}
Here $\Omega$ are natural frequencies distributed according to a unimodal density $g(\Omega)$,
and $P(\varphi|\Omega)$ is the probability density of oscillators
with natural frequency $\Omega$.

The theory of synchronization, developed by Kuramoto~\cite{Kuramoto1984}, predicts the
existence of a critical value of the coupling constant $\varepsilon$, beyond which
the macroscopic mean field and the frequency of the global phase assume constant 
values, $R>0$ and $\overline\Omega:=\dot\Phi$, respectively.
Oscillators with natural frequencies which satisfy $|\Omega-\overline\Omega|<\varepsilon R$
 are locked by the mean field, i.~e. rotate with $\bar\Omega$,
and constitute the ordered part of the population. 
Oscillators at the tails of the natural frequency distribution
with $|\Omega-\overline\Omega|>\varepsilon R$ each rotate with a different average frequency and constitute the disordered part.

To quantify order and disorder in the system, we calculate the 
pairwise cross-correlations between the oscillators.
First, we perform a shift to the mean field reference frame $\tilde\varphi := \varphi-\Phi$.
In the ordered part, the oscillators have constant phases $\tilde\varphi$ and are thus 
perfectly correlated. To
calculate cross-correlations in the disordered part, 
the phases $\tilde\varphi$ cannot be used directly, because
their probability density on the circle is not uniform: it is proportional 
to $\tilde P(\tilde\varphi|\Omega)
\sim|\Omega-\overline\Omega-\varepsilon R\sin\tilde\varphi|^{-1}$ \cite{pikovsky2003synchronization},
i.e.,~is a wrapped Cauchy distribution. This distribution is fully 
characterized by the first harmonic 
$z=\langle e^{i\tilde\varphi}\rangle=iq(1-\sqrt{1-q^{-2}})$, 
where $q=(\Omega-\overline\Omega)/(\varepsilon R)$ and 
$\langle\cdot\rangle$ is the average over oscillator phases $\tilde\varphi$ with density 
$\tilde P(\tilde\varphi|\Omega)$. 
The phases $\tilde\varphi$ can be transformed to uniformly distributed phase variables
$\psi$ by virtue of a M\"obius 
transform~\footnote{The transformation $\tilde\varphi\to\psi$ is an example 
of the protophase to phase transformation
used in the data analysis of oscillatory systems~\cite{Kralemann-08}.}
\begin{equation}
\exp[i\psi]=(\exp[i\tilde\varphi]-z)\big/(1-z^*\exp[i\tilde\varphi])~.
\label{eq:mt}
\end{equation}

Straightforward calculations show that 
$\dot{\psi}=\nu=(\Omega-\overline\Omega)
\left(1-q^{-2}\right)^{1/2}$, where $\nu$ denotes the observed frequency of the oscillator. 
Because the transformed phases $\psi$ rotate uniformly, 
with their frequency $\nu$ now only depending on intrinsic frequency $\Omega$,
we can straightforwardly apply the synchronization index -- a measure 
for the cross-correlation of two phases~\cite{mardia2009directional} -- as
\begin{equation}
\gamma_{12}=\left|\langle\exp[i(\psi_2-\psi_1)]\rangle\right|~.
\label{eq:cf}
\end{equation}
For two phase variables uniformly 
rotating with different frequencies $\nu_1$ and $\nu_2$, 
this yields zero, thus the
pairwise cross-correlation vanishes exactly in the disordered domain, 
independently of how close the
parameters of the oscillators are. Conversely, this cross-correlation is exactly 1 in the order domain.

\subsection{Mean field with external fluctuations} 
Our next goal is to show that non-vanishing cross-correlations appear in the disordered part if the mean field
contains added external noise.
We consider a simple modification of the Kuramoto model~\eqref{eq:km}: 
$Z=R\exp[i\overline\Omega t] +\tilde\sigma[\xi_1(t)+i\xi_2(t)]$, with constants $R, \overline\Omega$ 
and Gaussian random processes $\xi_1(t), \xi_2(t)$ of noise strength 
 $\tilde \sigma$ with $\langle \xi_i(t)\xi_j(t')\rangle=2\delta_{ij}\delta(t-t')$. We assume
noise to be weak, so that
it does not significantly change the individual 
statistical properties of the oscillators (the distribution remains a 
wrapped Cauchy distribution, see Ref.~\cite{Tyulkina_etal-18} for a quantification of small deviations
due to weak Gaussian noise). However, as we will see,
it induces cross-correlations in the disordered region. 
Performing the same transformation to obtain
 uniformly rotating phase variables $\psi$ as above, we obtain 
for the transformed phase variables $\psi$ a set of
Langevin equations with common noise terms $\eta_1, \eta_2$:
\begin{equation}
\dot\psi=\nu-(a+b\sin\psi)\eta_1(t)+c\,\cos\psi\,\eta_2(t)~.
\label{eq:lang}
\end{equation}
Here $\eta_1(t),\eta_2(t)$ are mutually uncorrelated Gaussian white noise forces common to all transformed phases $\psi$, 
$\langle \eta_i(t)\eta_j(t')\rangle=2\delta_{ij}\delta(t-t')$; 
$\nu=(\Omega-\overline\Omega)
\left(1-q^{-2}\right)^{1/2}$ is the observed frequency as above; parameters
$a=\varepsilon^2 R\tilde\sigma/\nu$, $b = \varepsilon\tilde\sigma(\Omega-\overline{\Omega})/\nu$, and
$c = \varepsilon\tilde\sigma$ are the effective noise strengths. 

We now consider two oscillators $\psi_1,\psi_2$ from the set~\eqref{eq:lang}.
We assume parameters of these oscillators to be close: $\nu_1=\nu+\rho/2$
and $\nu_2=\nu-\rho/2$, with $\rho\ll \nu$; and similar for $a_{1,2}$ and $b_{1,2}$.
To calculate the cross-correlation function~\eqref{eq:cf}, we first write,
starting from~\eqref{eq:lang}, the
Langevin equations for the difference $\alpha = \psi_1-\psi_2$ and 
the sum $\beta=\psi_1+\psi_2$ of the transformed phase variables:

\begin{align*}
\dot\alpha=&\rho-(a_1-a_2)\eta_1(t)-(b_1+b_2)\,\sin\frac{\alpha}{2}\cos\frac{\beta}{2}\, \eta_1(t)\\&
-(b_1-b_2)\cos\frac{\alpha}{2}\sin\frac{\beta}{2}\, \eta_1(t)-2c\sin\frac{\alpha}{2}\sin\frac{\beta}{2}\, \eta_2(t)\;,\\
\dot\beta=&2\nu - (a_1+a_2)\eta_1(t) - (b_1+b_2)\,\cos\frac{\alpha}{2}\sin\frac{\beta}{2}\, \eta_1(t)\\&-
(b_1-b_2)\sin\frac{\alpha}{2}\cos\frac{\beta}{2}\, \eta_1(t)+
2c\cos\frac{\alpha}{2}\cos\frac{\beta}{2}\, \eta_2(t).
\end{align*}

This system yields a Fokker-Plank equation for the density $W(\alpha,\beta,t)$, 
which, by virtue of averaging over the fast rotating variable $\beta$ with the method of multiple scales \cite{nayfeh1981introduction}, can be reduced to the 
following equation for the density of the phase difference $w(\alpha,t)$:
\begin{equation}
\begin{gathered}
\frac{\partial }{\partial t}w+\rho\frac{\partial}{\partial\alpha}w-
\left(\frac{(b_1+b_2)^2}{4}+c^2\right)\frac{\partial^2}{\partial\alpha^2}[(1-\cos\alpha)w]=\\
\frac{(b_1-b_2)^2}{4} \frac{\partial^2}{\partial\alpha^2}[(1+\cos\alpha)w]+
(a_1-a_2)^2\frac{\partial^2}{\partial\alpha^2} w\;.
\end{gathered}
\end{equation}
The terms on the r.h.s. of this equation are of second order in the small parameter $\rho$, and therefore
we can neglect them. As a result,
only the weighted sum of noise terms
$\sigma^2:=(b_1+b_2)^2/4+c^2 \approx \varepsilon^2\tilde\sigma^2\{1+[(\Omega-\overline{\Omega})/\nu]^2\}$
is relevant. 

The stationary solution of this equation can be straightforwardly written 
as an integral; the calculation of
the cross-correlation $\gamma=\left|\langle \E^{i\alpha}\rangle\right| = |\int_{-\pi}^\pi\,w(\alpha)\,\exp(i\alpha)\, \mathrm{d}\alpha|$ reduces to a nontrivial 
integration, which nevertheless can be expressed explicitly:
\begin{equation}
\begin{gathered}
\gamma^2=1+4d^2[\operatorname{ci}^2(2d)+\operatorname{si}^2(2d)]\\
-4d[\operatorname{ci}(2d)\sin(2d)-\operatorname{si}(2d)\cos(2d)]\;,
\end{gathered}
\label{eq:grho}
\end{equation}
where $\operatorname{ci}$ and $\operatorname{si}$ are the cosine and sine integral functions, 
and the ratio $d$ between the frequency mismatch $\rho$ and the noise strength $\sigma^2$, $d=\rho/\sigma^2$,
is the only parameter.
This cross-correlation function (cf. Fig.~\ref{fig:dep_on_a}, black solid line)
 tends to 1 for $d\ll 1$ 
and decays as $\gamma\sim 1/d$ 
as $d\gg 1$. Thus, our main analytical result~\eqref{eq:grho} 
shows that common external noise added to the mean field induces
cross-correlations in the disordered domain, with a characteristic 
cross-correlation length proportional to the noise intensity $\rho\sim \sigma^2$.

The physical explanation of this cross-correlation lies in the stabilizing effect of the common noise:
its action on an oscillator leads to a negative Lyapunov exponent, 
which results in complete synchronization of
identical oscillators~\cite{Goldobin-Pikovsky-04,Goldobin-Pikovsky-05b}. For nonidentical 
oscillators, the difference in the natural
frequencies prevents complete synchrony, but the phases are most of the time kept close
to each other by noise, with occasional fast phase slips~\cite{Goldobin2017} 
that account for the observed frequency difference.

\section{Finite-size mean field fluctuation} \label{sec:setup2}
As demonstrated above, micro-scale 
cross-correlations appear in the disordered domain
of the Kuramoto model in the thermodynamic limit with external mean field noise.
A natural question arises, if also the intrinsic order parameter fluctuations in deterministic finite ensembles generate such micro-scale cross-correlations. 
The essential parameter here is ensemble size $N$.
Simple estimations based on the theory above show
that the cross-correlations are rather small between typical pairs oscillators:
For an ensemble of size $N$, the
characteristic frequency mismatch between the oscillators is $\rho\sim N^{-1}$. However, in 
order to create sizeable fluctuations of the mean field, $N$ must be small. 
If one assumes $\sigma^2\sim N^{-1}$, then a \textit{typical} value of the
parameter $d = \rho/\sigma^2$ will be close to unity, which is too 
large for the cross-correlations to be observable (see Fig.~\ref{fig:dep_on_a}).

\subsection{A model with active and passive oscillators}

The size of the ensemble dictates not only the size of the intrinsic fluctuations of the
mean field, which tends to have an organizing effect on the phases, 
it also determines the \textit{typical} natural frequency mismatch of a given pair of oscillators, which
tends to have an disorganizing effect on their phases.
According to~\eqref{eq:grho}, the cross-correlation between the phases
with typical natural frequency difference for typical mean field fluctuation intensity (both depending on $N$) are small. The probability to find a pair with much smaller than average natural frequency difference is high, but then again it is difficult to disentangle different effects on such singular pairs. However, this
problem can be resolved if the fluctuation level (or the effective noise strength)
is decoupled from the range of natural frequency differences.

To decouple the two opposing effects, we introduce a modification of the Kuramoto model, where the oscillators are of twoc)
types: active ones $\phi_j$ ($j = 1, 2, \ldots, N$) with natural frequencies $\Omega^A_j$, 
and passive ones (tracers) $\varphi_k$ ($k = 1, 2, \ldots, M$) with natural frequencies $\Omega_k$. 
The oscillators of both types obey the same equation~\eqref{eq:km}. However, only the active oscillators 
contribute to the mean field:
$Z=R\exp[i\Phi]=N^{-1}\sum_{j=1}^N \exp[i\phi_j]$. Here $N$ is 
the number of active oscillators, while
the number of passive ones $M$ can be arbitrary (and they can have any 
distribution of frequencies). One can say that
passive oscillators ``test'' the mean field created by active 
oscillators, similar to how ideal fluid tracers ``test'' the flow of a fluid.
The passive oscillators do so at different frequencies, especially
at those not presented in the active set. Similar technique
has been used in~\cite{Rosenblum-Pikovsky-etal-02}
to determine the frequency of chaotic signals via locking.

Equivalently, the system of active-passive oscillators can be considered as a large network
\begin{equation}
\dot \varphi_k=\Omega_k+\frac{\varepsilon}{N}\sum_j K_{kj}\sin(\varphi_j-\varphi_k)\;,
\label{eq:acpas}
\end{equation}
 where
$K_{kj}=1$ if the phase $\varphi_j$ belongs to the active set, and $K_{kj}=0$ otherwise.
Experimentally, such a coupling has been directly implemented in a set of 2816 optically
coupled periodic chemical Belousov-Zhabotinsky reactors~\cite{Totz_etal-17}.

Similar setups are often used in systems with long-range interactions, 
for example, in restricted $N$-body problems in gravitational 
systems. Heavy bodies such as planets, stars and galaxies contribute to the gravitational
field in which they move, while other, lighter particles move in the same field,
but their contribution to the field is negligible. In a more general context of interdisciplinary applications
of complex systems, the division into active and passive agents occurs by itself
in macro-social opinion formation processes. In social media,
a few forward thinkers (or influencers) lead the public discourse by writing texts and comments,
while the opinions of a large number of passive users (followers) 
remain hidden: they follow the discussion without contributing to it
~\cite{Gerson_etal-17}.

\subsection{Fluctuations beyond the synchronization transition}
As we show below, using tracers,
the micro-scale cross-correlations (and other
interesting features) and can be easily detected.
In this subsection, we illustrate these features for
a partially synchronized state of the finite-size Kuramoto model,
and in subsection~\ref{sec:fbel} for a
state below the synchronization transition.

\begin{figure}
\includegraphics[width = 0.45\textwidth]{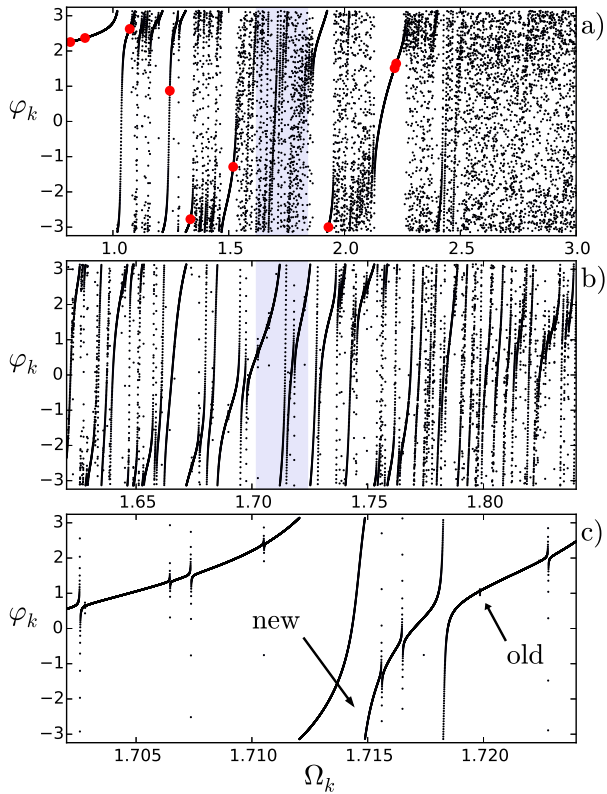}
\caption{Snapshot of passive oscillators (black dots) and active oscillators 
(large red dots) phases plotted against 
natural frequencies for a finite ensemble of $50$ active oscillators (not all shown).
The mean field is created by the active oscillators with natural 
frequencies drawn randomly from a Gaussian distribution (same sample as in 
Fig.~\ref{fig:gamma_50_obs}) at slightly super-critical 
coupling $\varepsilon = 1.85$ (same in Fig.~\ref{fig:gamma_50_obs}). Three zoom levels 
of factor $10$ are marked by shaded areas. Arrows in c) 
point at $\Omega_k \approx 1.7145$ and $\Omega_k \approx 1.72$ 
for a new and an old phase slips, respectively. 
See Supplementary Material at [URL will be inserted by publisher] for an animated version of this figure. One second in the video equals one time unit. 
}
\label{fig:bars}
\end{figure}

First, we give a qualitative picture
of the cross-correlations. Fig.~\ref{fig:bars} shows a snapshot of phases
for a population of $N=50$ active oscillators together with 
a set of $M=5\cdot10^4$ tracers. The natural frequencies of the active ones
are sampled from a normal distribution with zero mean and unit variance, 
for which the critical
coupling constant in the thermodynamic limit is $\varepsilon_c\approx 1.6$.
The micro-scale ordering effect becomes evident
if one zooms in to increasingly smaller scales, from panel a) to panel c). Panel c) shows the 
characteristic correlated state
profile of the tracers'  phases, consisting of ruptured nearly horizontal bars. A bar is formed due to
the ordering action of the fluctuations of the mean field, which synchronize passive oscillators 
with close frequencies.
Ruptures appear when  oscillators with higher frequencies make an additional
rotation (a phase slip) with respect to oscillators with  smaller (but similar) frequencies. In Fig.~\ref{fig:bars}c)
one can clearly see a fresh phase slip around $\Omega\approx 1.7145$, an older less pronounced
phase slip around $\Omega\approx 1.717$, and several old phase slips that have 
almost disappeared. The phase slips 
become less visible over time because of the stabilizing effect reflected in a negative 
Lyapunov exponent as outlined above.

The micro-correlated structures like Fig.~\ref{fig:bars}c) are observed in all disordered domains
visible in the global picture Fig.~\ref{fig:bars}a). Additionally,
macroscopically ordered regions are seen close to the active oscillators
not entrained by the mean field (with $\Omega_k\gtrsim 1.0$) (Fig.~\ref{fig:bars}a).
Here the tracers are synchronized to the active units 
(like the satellites are trapped by their planet's gravitational field).
This is because the fluctuations of the mean field are in fact not completely random, but
contain relatively strong nearly periodic components from the non-entrained active 
oscillators. These components suffice,
at least for small ensembles, 
to fully entrain tracers with natural frequencies close to a common active oscillator.

\begin{figure}
 \centering
 \setlength{\unitlength}{1cm}
 \begin{picture}(8,11.3)
\put(0,0){ \includegraphics[width = \columnwidth]{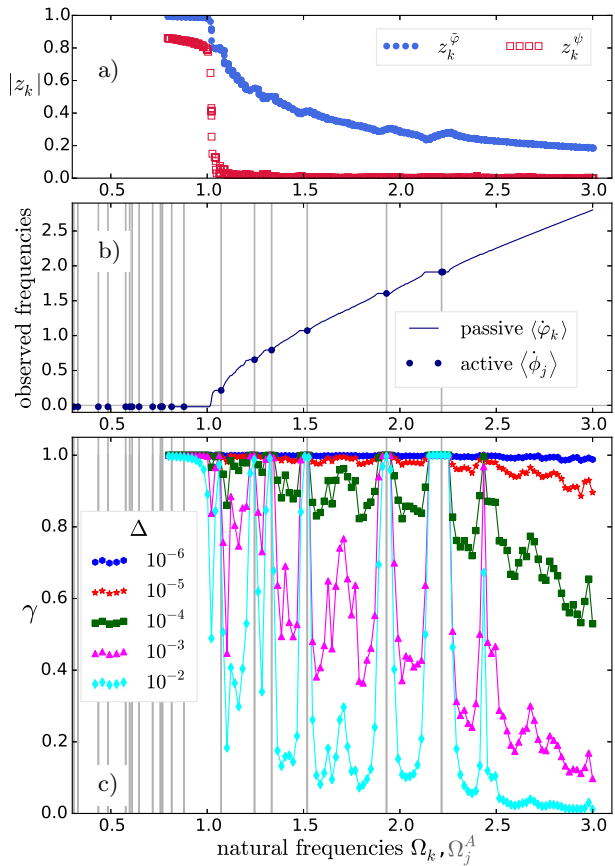}}
\put(1.3,10.5){\colorbox{white}{a)}}
\put(1.3,8.1){\colorbox{white}{b)}}
\put(1.3,1){\colorbox{white}{c)}}
\end{picture}
 \caption{Observed cross-correlation as defined by Eq.~\eqref{eq:cf} of passive oscillators 
 coupled to the mean field $Z$ of $50$ active oscillators against their natural frequencies (mean $\langle R\rangle = 0.58$, variance $\sigma_R^2 =3.5\cdot10^{-3}$),
 shown for various levels of frequency mismatch $\Delta$ between passive oscillators.
 a): Absolute value of the (time-averaged) first harmonic
 of individual passive oscillators 
 before ($z$, blue circles) and 
 after ($z_\psi$, red squares) the M\"{o}bius transform.
 $z_\psi$ drops to near zero after the transformation,
 which shows that $\psi$ is now uniformly rotating.
 b): Observed vs. natural frequencies of passive (line) 
 and active oscillators (dots). 
 Gray vertical lines in  panels b), c) mark the natural 
 frequencies of the active oscillators. c): Synchronization index $\gamma$ 
 between pairs of passive oscillators with  natural 
 frequencies $\Omega_k\pm\Delta/2$.}
 \label{fig:gamma_50_obs}
\end{figure}

Now we quantify the cross-correlations illustrated in Fig.~\ref{fig:bars}. 
We calculate the cross-correlation coefficient 
$\gamma(\Omega,\Delta)$ for two tracers with natural frequencies $\Omega-\Delta/2,\,
\Omega+\Delta/2$
according to \eqref{eq:cf}. To this end, we need to perform the M\"obius 
transformation~\eqref{eq:mt} to
obtain uniformly distributed phase variables $\psi$. First, we calculate the time dependent
difference between the tracer phases and the mean field phase 
$\tilde \varphi(t)=\varphi(t)-\Phi(t)$.
We then average these phases over time, $z=\langle \exp[i\tilde\varphi]\rangle$, 
which gives the empirical value of the
parameter characterizing the wrapped Cauchy distribution of 
$\tilde\varphi$. Then, the M\"obius 
transform~\eqref{eq:mt} is applied. To check that we indeed obtained the uniformly 
distributed 
phase variable $\psi$, we calculate the first harmonics 
$z_\psi=\langle \exp[i\psi]\rangle$ 
and compare it to $z$ (Fig.~\ref{fig:gamma_50_obs}a): 
one can see that indeed the transformation
yields a uniformly distributed set of phase variables
(within reasonable tolerance), because the amplitudes of the 
time averages of their first harmonics become very close to zero 
after the transformation.

In Fig.~\ref{fig:gamma_50_obs} b) we show the observed frequencies 
of the tracers and the active oscillators.
One can clearly see synchronized neighborhoods of active units as 
plateaus in this graph. Outside the plateaus,
the tracers are not locked and their observed frequency varies 
continuously with their natural one. It is in these
domains outside the plateaus, that the micro-scale cross-correlations 
can be observed and measured, as
illustrated in Fig.~\ref{fig:gamma_50_obs} c). Here we show 
values of the cross-correlation coefficient
$\gamma(\Omega,\Delta)$ for several values of frequency mismatch $\Delta$: cross-correlation is nearly
perfect for $\Delta\lesssim 10^{-5}$, while for $\Delta\gtrsim 0.01$ the values
of the coefficient typically do not exceed $0.5$.

 \subsection{Fluctuations below the synchronization transition}
 \label{sec:fbel}
 
\begin{figure}
 \centering
 \setlength{\unitlength}{1cm}
 \begin{picture}(8,6.1)
 \put(-.4,0){  \includegraphics[width = \columnwidth]{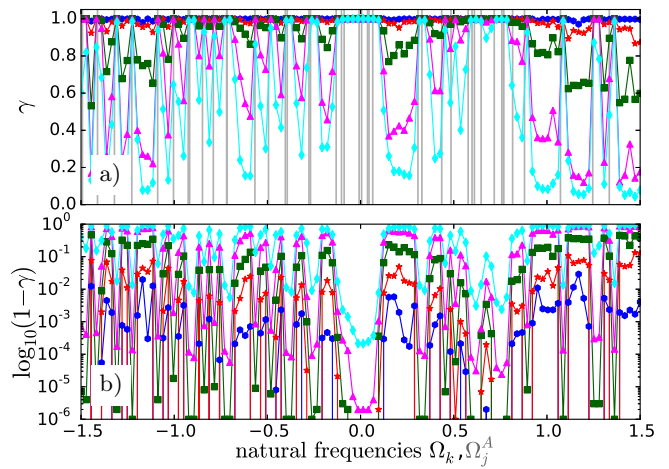}}
  \put(0.8,3.8){\colorbox{white}{a)}}
\put(0.8,1.){\colorbox{white}{b)}}
 \end{picture}
 
\caption{The cross-correlations of passive oscillators coupled to a mean field of $50$ active ones at $\varepsilon=1$ (same natural frequency sample as Fig.~\ref{fig:gamma_50_obs}). The mean field $Z$ fluctuates around zero, see~\cite{FPAP2018}. Panel a):
cross-correlations in the linear scale; panel b): the same in logarithmic scale to resolve the
region $\gamma\lesssim 1$. Markers and colors as in Fig.~\ref{fig:gamma_50_obs} c).}
\label{fig:fig4}
\end{figure}

 To show that the effect of microscopic cross-correlations occurs for 
subcritical values of coupling constant $\varepsilon$ as well, we present the
calculations of the cross-correlations for an ensemble of $N=50$ oscillators
with natural frequencies randomly sampled from the standard normal distribution, and
$\varepsilon=1$ in Fig.~\ref{fig:fig4}. For this relatively small coupling, the complex mean field
fluctuates around zero, see e.~g.~\cite{FPAP2018}. Therefore, a
transformation of the phases to uniformly rotating ones ($\tilde\varphi\to\psi$) is unnecessary, contrary
to the case of a non-vanishing mean field at stronger coupling (Figs.~\ref{fig:bars}, ~\ref{fig:gamma_50_obs}).
For better visibility of both high and low cross-correlations, we present in
Fig.~\ref{fig:fig4} the cross-correlation constant $\gamma(\Omega,\Delta)$
in linear and logarithmic scales.
The multiple locked regions relate to the frequencies
of active oscillators. The central region around $\Omega\approx 0$ corresponds
to the frequency of a synchronous cluster that has already been formed by the
oscillators at the center of the locking region, 
even though this cluster is still not large enough to ensure the existence
of a macroscopic mean field.
The figure shows that the microscopic cross-correlations are of a universal nature
and can be observed both below and above the synchronization transition.
 
\begin{figure}
 \psfrag{xlab}[cc][cc]{$\log_{10}(d)$}
 \psfrag{ylab}[cc][cc]{$\gamma(d)$}
 \includegraphics[width = \columnwidth]
 {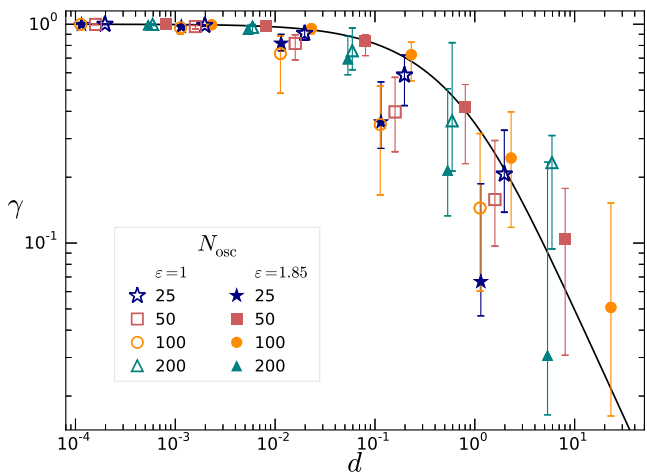}
\caption{Comparison of the observed cross-correlation $\gamma$ 
in sub- and super-critical ensembles ($\varepsilon = 1$ and $1.85$ with open and filled symbols, respectively) of different sizes $N_{osc}= 25, 50, 100, 200$ with the theoretical expression~\eqref{eq:grho} (solid line).
The data points were generated in experiments as in Figs.~\ref{fig:gamma_50_obs},~\ref{fig:fig4}. The median of all $\gamma$'s for which $\gamma(10^{-2})<0.5$ (thereby excluding locked passive oscillators) is represented with error bars that mark the 25th and 75th percentile, respectively. Value $d=\Delta/\sigma^2$ is determined from the average over $\sigma^2 = 2\varepsilon^2 D\{1+(|\Omega_k-\bar\Omega|/\Delta)^2\}$ with diffusion coefficient $D$ that was integrated from the auto-correlation function of $R$. For each ensemble size, only one set of natural frequencies is represented.
}
\label{fig:dep_on_a}
\end{figure}

Finally, we illustrate in Fig.~\ref{fig:dep_on_a} 
a dependence of the cross-correlations on the noise level.
Unfortunately, a quantitative comparison with the theoretical prediction
\eqref{eq:grho} is not possible because the intrinsic fluctuations 
due to the finite-sized effect are very far
from being delta-correlated, as is assumed in the analytical theory. 
Nevertheless, for a qualitative comparison,
we calculated the autocorrelation function of the mean field, which has a peak at zero and 
pronounced oscillations due to the nearly periodic contributions of particular oscillators.
As a measure of the effective noise
intensity we took the difference $S$ between the peak value and the next largest maximum
of the cross-correlations.
This quantity decreases with the 
system size of the active ensemble. Furthermore,
we have chosen only non-locked passive units. One can see that the scaling 
relation $\gamma=\gamma(\Delta/S)$
follows at least qualitatively the theoretical curve \eqref{eq:grho},
although  a huge diversity of the observed cross-correlations,
due to the ``coloredness'' of the intrinsic finite-size mean field fluctuations, is also evident.
 
 \section{Conclusion} 
 In summary, we have shown that the fluctuations of the mean field in the Kuramoto model,
either externally imposed on the ensemble of infinite size, or naturally induced in the finite-sized model, 
lead to the appearance of cross-correlations in the
 disordered part of oscillator populations. These cross-correlations result
 from the competition between synchronization by common noise and desynchronization
 due to the parameter differences (usually the differences in the natural frequencies).  We have developed an analytical
 theory of these cross-correlations for a mean field being a constant (in a properly rotating frame)
 plus additional Gaussian white noise, 
 summarized in expression~\eqref{eq:grho}. 
  This theoretical result is directly applicable to models
 similar to the Kuramoto model, e.g., to the Kuramoto-Sakaguchi model,
where the mean field of a population is subject to external
fluctuations. In the derivation of \eqref{eq:grho} we explicitly
restrict the mean-field coupling to the first harmonics of the oscillator phases only. 
The case of a more general Daido-type coupling function requires extra analysis,
although qualitative arguments imply that the cross-correlations will be observed there as well. 

Furthermore, we have numerically characterized pairwise cross-correlations between passive oscillators
driven by the intrinsically fluctuating mean fields in a finite ensemble at two different coupling strengths, super- and sub-critical, respectively. In both cases,
the mean field contains nearly periodic components, therefore there exist locked regions
where high values of pairwise cross-correlations arise due to 
resonant locking. Between these locked regions,
the cross-correlations decay with increasing frequency mismatch, which is
in a qualitative agreement with the theory based on white noise.
The rather good quantitative agreement between white noise theory and finite Kuramoto model surprises, as, due to the dominant nearly periodic components
in the fluctuating mean field, fluctuations in the finite Kuramoto model are far from being ``white''. 

Overall, the effect is expected to be most pronounced 
 in situations where finite-size fluctuations are anomalously large  
 (populations with equidistant natural frequencies at sub-critical 
 coupling appear, as our preliminary
 calculations show, to belong to this class).
 
 We expect that this phenomenon is not restricted to the
 mean-field coupling, and can be observed in other large systems where 
 synchronized and disordered sub-populations coexist. 
 A prominent example here is a chimera state in a 
 one- or two-dimensional oscillatory medium with long-range 
 interactions~\cite{Kuramoto-Battogtokh-02}.
 A population of oscillators driven by two mean fields~\cite{Zhang_Pikovsky_Liu_17} 
 also demonstrated nontrivial regimes with a coexistence of ordered and disordered subpopulations.
 Indeed, chimera states
 are defined as the coexistence of coherent and non-coherent domains among 
 identical oscillators, and finite-size 
 fluctuations~\cite{Wolfrum-Omelchenko-11} may lead to 
 cross-correlations in the disordered domain;
 this issue is currently under consideration.
 

\acknowledgments
This paper was developed within the scope of the IRTG 1740 / TRP 2015/50122-0, 
funded by the DFG/ FAPESP. A.~P.~was supported by 
the Russian Science Foundation (grant No.\ 17-12-01534). 
The authors thank D.~Goldobin, R.~Cestnik, M.~Wolfrum, O.~Omel'chenko, D.~Paz\'{o}, H.~Engel,
and R.~Toenjes for helpful discussions.


\begin{thebibliography}{31}%
\makeatletter
\providecommand \@ifxundefined [1]{%
 \@ifx{#1\undefined}
}%
\providecommand \@ifnum [1]{%
 \ifnum #1\expandafter \@firstoftwo
 \else \expandafter \@secondoftwo
 \fi
}%
\providecommand \@ifx [1]{%
 \ifx #1\expandafter \@firstoftwo
 \else \expandafter \@secondoftwo
 \fi
}%
\providecommand \natexlab [1]{#1}%
\providecommand \enquote  [1]{``#1''}%
\providecommand \bibnamefont  [1]{#1}%
\providecommand \bibfnamefont [1]{#1}%
\providecommand \citenamefont [1]{#1}%
\providecommand \href@noop [0]{\@secondoftwo}%
\providecommand \href [0]{\begingroup \@sanitize@url \@href}%
\providecommand \@href[1]{\@@startlink{#1}\@@href}%
\providecommand \@@href[1]{\endgroup#1\@@endlink}%
\providecommand \@sanitize@url [0]{\catcode `\\12\catcode `\$12\catcode
  `\&12\catcode `\#12\catcode `\^12\catcode `\_12\catcode `\%12\relax}%
\providecommand \@@startlink[1]{}%
\providecommand \@@endlink[0]{}%
\providecommand \url  [0]{\begingroup\@sanitize@url \@url }%
\providecommand \@url [1]{\endgroup\@href {#1}{\urlprefix }}%
\providecommand \urlprefix  [0]{URL }%
\providecommand \Eprint [0]{\href }%
\providecommand \doibase [0]{http://dx.doi.org/}%
\providecommand \selectlanguage [0]{\@gobble}%
\providecommand \bibinfo  [0]{\@secondoftwo}%
\providecommand \bibfield  [0]{\@secondoftwo}%
\providecommand \translation [1]{[#1]}%
\providecommand \BibitemOpen [0]{}%
\providecommand \bibitemStop [0]{}%
\providecommand \bibitemNoStop [0]{.\EOS\space}%
\providecommand \EOS [0]{\spacefactor3000\relax}%
\providecommand \BibitemShut  [1]{\csname bibitem#1\endcsname}%
\let\auto@bib@innerbib\@empty
\bibitem [{\citenamefont {Kuramoto}(1984)}]{Kuramoto1984}%
  \BibitemOpen
  \bibfield  {author} {\bibinfo {author} {\bibfnamefont {Y.}~\bibnamefont
  {Kuramoto}},\ }\href@noop {} {\emph {\bibinfo {title} {{Chemical
  Oscillations, Waves, and Turbulence}}}},\ \bibinfo {edition} {{S}pringer
  {S}eries in {S}ynergetics}\ ed.,\ Vol.~\bibinfo {volume} {19}\ (\bibinfo
  {publisher} {Springer Berlin Heidelberg},\ \bibinfo {year}
  {1984})\BibitemShut {NoStop}%
\bibitem [{\citenamefont {Strogatz}(2003)}]{Strogatz-03}%
  \BibitemOpen
  \bibfield  {author} {\bibinfo {author} {\bibfnamefont {S.~H.}\ \bibnamefont
  {Strogatz}},\ }\href@noop {} {\emph {\bibinfo {title} {Sync: {T}he Emerging
  Science of Spontaneous Order}}}\ (\bibinfo  {publisher} {Hyperion},\ \bibinfo
  {address} {NY},\ \bibinfo {year} {2003})\BibitemShut {NoStop}%
\bibitem [{\citenamefont {Pikovsky}\ \emph {et~al.}(2001)\citenamefont
  {Pikovsky}, \citenamefont {Rosenblum},\ and\ \citenamefont
  {Kurths}}]{pikovsky2003synchronization}%
  \BibitemOpen
  \bibfield  {author} {\bibinfo {author} {\bibfnamefont {A.}~\bibnamefont
  {Pikovsky}}, \bibinfo {author} {\bibfnamefont {M.}~\bibnamefont {Rosenblum}},
  \ and\ \bibinfo {author} {\bibfnamefont {J.}~\bibnamefont {Kurths}},\
  }\href@noop {} {\emph {\bibinfo {title} {Synchronization: A Universal Concept
  in Nonlinear Sciences}}}\ (\bibinfo  {publisher} {Cambridge University
  Press},\ \bibinfo {year} {2001})\BibitemShut {NoStop}%
\bibitem [{\citenamefont {Acebr{\'o}n}\ \emph {et~al.}(2005)\citenamefont
  {Acebr{\'o}n}, \citenamefont {Bonilla}, \citenamefont {Vicente},
  \citenamefont {Ritort},\ and\ \citenamefont {Spigler}}]{Acebron-etal-05}%
  \BibitemOpen
  \bibfield  {author} {\bibinfo {author} {\bibfnamefont {J.~A.}\ \bibnamefont
  {Acebr{\'o}n}}, \bibinfo {author} {\bibfnamefont {L.~L.}\ \bibnamefont
  {Bonilla}}, \bibinfo {author} {\bibfnamefont {C.~J.~P.}\ \bibnamefont
  {Vicente}}, \bibinfo {author} {\bibfnamefont {F.}~\bibnamefont {Ritort}}, \
  and\ \bibinfo {author} {\bibfnamefont {R.}~\bibnamefont {Spigler}},\
  }\href@noop {} {\bibfield  {journal} {\bibinfo  {journal} {Rev. Mod. Phys.}\
  }\textbf {\bibinfo {volume} {77}},\ \bibinfo {pages} {137} (\bibinfo {year}
  {2005})}\BibitemShut {NoStop}%
\bibitem [{\citenamefont {Ott}\ and\ \citenamefont {Antonsen}(2008)}]{Ott2008}%
  \BibitemOpen
  \bibfield  {author} {\bibinfo {author} {\bibfnamefont {E.}~\bibnamefont
  {Ott}}\ and\ \bibinfo {author} {\bibfnamefont {T.~M.}\ \bibnamefont
  {Antonsen}},\ }\href@noop {} {\bibfield  {journal} {\bibinfo  {journal}
  {Chaos}\ }\textbf {\bibinfo {volume} {18}} (\bibinfo {year}
  {2008})}\BibitemShut {NoStop}%
\bibitem [{\citenamefont {Daido}(1987)}]{Daido1987}%
  \BibitemOpen
  \bibfield  {author} {\bibinfo {author} {\bibfnamefont {H.}~\bibnamefont
  {Daido}},\ }\href@noop {} {\bibfield  {journal} {\bibinfo  {journal} {Journal
  of Physics A: Mathematical and General}\ }\textbf {\bibinfo {volume} {20}},\
  \bibinfo {pages} {L629} (\bibinfo {year} {1987})}\BibitemShut {NoStop}%
\bibitem [{\citenamefont {Hong}\ \emph {et~al.}(2007)\citenamefont {Hong},
  \citenamefont {Chat{\'{e}}}, \citenamefont {Park},\ and\ \citenamefont
  {Tang}}]{Hong2007}%
  \BibitemOpen
  \bibfield  {author} {\bibinfo {author} {\bibfnamefont {H.}~\bibnamefont
  {Hong}}, \bibinfo {author} {\bibfnamefont {H.}~\bibnamefont {Chat{\'{e}}}},
  \bibinfo {author} {\bibfnamefont {H.}~\bibnamefont {Park}}, \ and\ \bibinfo
  {author} {\bibfnamefont {L.~H.}\ \bibnamefont {Tang}},\ }\href@noop {}
  {\bibfield  {journal} {\bibinfo  {journal} {Phys. Rev. Lett.}\ }\textbf
  {\bibinfo {volume} {99}},\ \bibinfo {pages} {1} (\bibinfo {year}
  {2007})}\BibitemShut {NoStop}%
\bibitem [{\citenamefont {Nishikawa}\ \emph {et~al.}(2014)\citenamefont
  {Nishikawa}, \citenamefont {Iwayama}, \citenamefont {Tanaka}, \citenamefont
  {Horita},\ and\ \citenamefont {Aihara}}]{Nishikawa2014}%
  \BibitemOpen
  \bibfield  {author} {\bibinfo {author} {\bibfnamefont {I.}~\bibnamefont
  {Nishikawa}}, \bibinfo {author} {\bibfnamefont {K.}~\bibnamefont {Iwayama}},
  \bibinfo {author} {\bibfnamefont {G.}~\bibnamefont {Tanaka}}, \bibinfo
  {author} {\bibfnamefont {T.}~\bibnamefont {Horita}}, \ and\ \bibinfo {author}
  {\bibfnamefont {K.}~\bibnamefont {Aihara}},\ }\href@noop {} {\bibfield
  {journal} {\bibinfo  {journal} {Progress of Theoretical and Experimental
  Physics}\ }\textbf {\bibinfo {volume} {2014}},\ \bibinfo {pages} {1}
  (\bibinfo {year} {2014})}\BibitemShut {NoStop}%
\bibitem [{\citenamefont {Hong}\ \emph {et~al.}(2015)\citenamefont {Hong},
  \citenamefont {Chat\'e}, \citenamefont {Tang},\ and\ \citenamefont
  {Park}}]{Hong_etal-15}%
  \BibitemOpen
  \bibfield  {author} {\bibinfo {author} {\bibfnamefont {H.}~\bibnamefont
  {Hong}}, \bibinfo {author} {\bibfnamefont {H.}~\bibnamefont {Chat\'e}},
  \bibinfo {author} {\bibfnamefont {L.-H.}\ \bibnamefont {Tang}}, \ and\
  \bibinfo {author} {\bibfnamefont {H.}~\bibnamefont {Park}},\ }\href@noop {}
  {\bibfield  {journal} {\bibinfo  {journal} {Phys. Rev. E}\ }\textbf {\bibinfo
  {volume} {92}},\ \bibinfo {pages} {022122} (\bibinfo {year}
  {2015})}\BibitemShut {NoStop}%
\bibitem [{\citenamefont {Peter}\ and\ \citenamefont
  {Pikovsky}(2018)}]{FPAP2018}%
  \BibitemOpen
  \bibfield  {author} {\bibinfo {author} {\bibfnamefont {F.}~\bibnamefont
  {Peter}}\ and\ \bibinfo {author} {\bibfnamefont {A.}~\bibnamefont
  {Pikovsky}},\ }\href@noop {} {\bibfield  {journal} {\bibinfo  {journal}
  {Phys. Rev. E}\ }\textbf {\bibinfo {volume} {97}},\ \bibinfo {pages} {032310}
  (\bibinfo {year} {2018})}\BibitemShut {NoStop}%
\bibitem [{\citenamefont {Pikovsky}\ and\ \citenamefont {Politi}(2016)}]{LE}%
  \BibitemOpen
  \bibfield  {author} {\bibinfo {author} {\bibfnamefont {A.}~\bibnamefont
  {Pikovsky}}\ and\ \bibinfo {author} {\bibfnamefont {A.}~\bibnamefont
  {Politi}},\ }\href@noop {} {\emph {\bibinfo {title} {{Lyapunov Exponents: A
  Tool to Explore Complex Dynamics}}}}\ (\bibinfo  {publisher} {Cambridge
  University Press},\ \bibinfo {year} {2016})\ Chap.~\bibinfo {chapter}
  {10}\BibitemShut {NoStop}%
\bibitem [{\citenamefont {Gilad}(2013)}]{Gilad2013}%
  \BibitemOpen
  \bibfield  {author} {\bibinfo {author} {\bibfnamefont {B.}~\bibnamefont
  {Gilad}},\ }\emph {\bibinfo {title} {Synchronization of Network Coupled
  Chaotic and Oscillatory Dynamical Systems}},\ \href@noop {} {Ph.D. thesis},\
  \bibinfo  {school} {University of Maryland} (\bibinfo {year}
  {2013})\BibitemShut {NoStop}%
\bibitem [{\citenamefont {Pikovsky}\ \emph {et~al.}(1994)\citenamefont
  {Pikovsky}, \citenamefont {Rateitschak},\ and\ \citenamefont
  {Kurths}}]{Pikovsky-Rateitschak-Kurths-94}%
  \BibitemOpen
  \bibfield  {author} {\bibinfo {author} {\bibfnamefont {A.~S.}\ \bibnamefont
  {Pikovsky}}, \bibinfo {author} {\bibfnamefont {K.}~\bibnamefont
  {Rateitschak}}, \ and\ \bibinfo {author} {\bibfnamefont {J.}~\bibnamefont
  {Kurths}},\ }\href@noop {} {\bibfield  {journal} {\bibinfo  {journal} {Z.
  Physik B}\ }\textbf {\bibinfo {volume} {95}},\ \bibinfo {pages} {541}
  (\bibinfo {year} {1994})}\BibitemShut {NoStop}%
\bibitem [{\citenamefont {Komarov}\ and\ \citenamefont
  {Pikovsky}(2015)}]{Komarov-Pikovsky-15b}%
  \BibitemOpen
  \bibfield  {author} {\bibinfo {author} {\bibfnamefont {M.}~\bibnamefont
  {Komarov}}\ and\ \bibinfo {author} {\bibfnamefont {A.}~\bibnamefont
  {Pikovsky}},\ }\href@noop {} {\bibfield  {journal} {\bibinfo  {journal}
  {Phys. Rev. E}\ }\textbf {\bibinfo {volume} {92}},\ \bibinfo {pages} {020901}
  (\bibinfo {year} {2015})}\BibitemShut {NoStop}%
\bibitem [{\citenamefont {Pikovsky}\ \emph {et~al.}(2002)\citenamefont
  {Pikovsky}, \citenamefont {Zaikin},\ and\ \citenamefont {de~la
  Casa}}]{Pikovsky-Zaikin-Casa-02}%
  \BibitemOpen
  \bibfield  {author} {\bibinfo {author} {\bibfnamefont {A.}~\bibnamefont
  {Pikovsky}}, \bibinfo {author} {\bibfnamefont {A.}~\bibnamefont {Zaikin}}, \
  and\ \bibinfo {author} {\bibfnamefont {M.~A.}\ \bibnamefont {de~la Casa}},\
  }\href@noop {} {\bibfield  {journal} {\bibinfo  {journal} {Phys. Rev. Lett.}\
  }\textbf {\bibinfo {volume} {88}},\ \bibinfo {pages} {050601} (\bibinfo
  {year} {2002})}\BibitemShut {NoStop}%
\bibitem [{\citenamefont {Goldobin}\ and\ \citenamefont
  {Pikovsky}(2004)}]{Goldobin-Pikovsky-04}%
  \BibitemOpen
  \bibfield  {author} {\bibinfo {author} {\bibfnamefont {D.~S.}\ \bibnamefont
  {Goldobin}}\ and\ \bibinfo {author} {\bibfnamefont {A.~S.}\ \bibnamefont
  {Pikovsky}},\ }\href@noop {} {\bibfield  {journal} {\bibinfo  {journal}
  {Radiophysics and Quantum Electronics}\ }\textbf {\bibinfo {volume} {47}},\
  \bibinfo {pages} {910} (\bibinfo {year} {2004})}\BibitemShut {NoStop}%
\bibitem [{\citenamefont {Goldobin}\ and\ \citenamefont
  {Pikovsky}(2005)}]{Goldobin-Pikovsky-05b}%
  \BibitemOpen
  \bibfield  {author} {\bibinfo {author} {\bibfnamefont {D.~S.}\ \bibnamefont
  {Goldobin}}\ and\ \bibinfo {author} {\bibfnamefont {A.}~\bibnamefont
  {Pikovsky}},\ }\href@noop {} {\bibfield  {journal} {\bibinfo  {journal}
  {Phys. Rev. E}\ }\textbf {\bibinfo {volume} {71}},\ \bibinfo {pages}
  {045201(R)} (\bibinfo {year} {2005})}\BibitemShut {NoStop}%
\bibitem [{\citenamefont {Nagai}\ and\ \citenamefont
  {Kori}(2010)}]{Nagai_Kori_2010}%
  \BibitemOpen
  \bibfield  {author} {\bibinfo {author} {\bibfnamefont {K.~H.}\ \bibnamefont
  {Nagai}}\ and\ \bibinfo {author} {\bibfnamefont {H.}~\bibnamefont {Kori}},\
  }\href@noop {} {\bibfield  {journal} {\bibinfo  {journal} {Phys. Rev. E}\
  }\textbf {\bibinfo {volume} {81}},\ \bibinfo {pages} {065202} (\bibinfo
  {year} {2010})}\BibitemShut {NoStop}%
\bibitem [{\citenamefont {Pimenova}\ \emph {et~al.}(2016)\citenamefont
  {Pimenova}, \citenamefont {Goldobin}, \citenamefont {Rosenblum},\ and\
  \citenamefont {Pikovsky}}]{Goldobin2016}%
  \BibitemOpen
  \bibfield  {author} {\bibinfo {author} {\bibfnamefont {A.~V.}\ \bibnamefont
  {Pimenova}}, \bibinfo {author} {\bibfnamefont {D.~S.}\ \bibnamefont
  {Goldobin}}, \bibinfo {author} {\bibfnamefont {M.}~\bibnamefont {Rosenblum}},
  \ and\ \bibinfo {author} {\bibfnamefont {A.}~\bibnamefont {Pikovsky}},\
  }\href@noop {} {\bibfield  {journal} {\bibinfo  {journal} {Scientific
  Reports}\ }\textbf {\bibinfo {volume} {6}},\ \bibinfo {pages} {38518}
  (\bibinfo {year} {2016})}\BibitemShut {NoStop}%
\bibitem [{Note1()}]{Note1}%
  \BibitemOpen
  \bibinfo {note} {The transformation $\protect \mathaccentV {tilde}07E\varphi
  \to \psi $ is an example of the protophase to phase transformation used in
  the data analysis of oscillatory systems~\cite {Kralemann-08}.}\BibitemShut
  {Stop}%
\bibitem [{\citenamefont {Mardia}\ and\ \citenamefont
  {Jupp}(2009)}]{mardia2009directional}%
  \BibitemOpen
  \bibfield  {author} {\bibinfo {author} {\bibfnamefont {K.}~\bibnamefont
  {Mardia}}\ and\ \bibinfo {author} {\bibfnamefont {P.}~\bibnamefont {Jupp}},\
  }\href@noop {} {\emph {\bibinfo {title} {Directional Statistics}}},\ Wiley
  Series in Probability and Statistics\ (\bibinfo  {publisher} {Wiley},\
  \bibinfo {year} {2009})\BibitemShut {NoStop}%
\bibitem [{\citenamefont {Tyulkina}\ \emph {et~al.}(2018)\citenamefont
  {Tyulkina}, \citenamefont {Goldobin}, \citenamefont {Klimenko},\ and\
  \citenamefont {Pikovsky}}]{Tyulkina_etal-18}%
  \BibitemOpen
  \bibfield  {author} {\bibinfo {author} {\bibfnamefont {I.~V.}\ \bibnamefont
  {Tyulkina}}, \bibinfo {author} {\bibfnamefont {D.~S.}\ \bibnamefont
  {Goldobin}}, \bibinfo {author} {\bibfnamefont {L.~S.}\ \bibnamefont
  {Klimenko}}, \ and\ \bibinfo {author} {\bibfnamefont {A.}~\bibnamefont
  {Pikovsky}},\ }\href@noop {} {\bibfield  {journal} {\bibinfo  {journal}
  {Phys. Rev. Lett.}\ }\textbf {\bibinfo {volume} {120}},\ \bibinfo {pages}
  {264101} (\bibinfo {year} {2018})}\BibitemShut {NoStop}%
\bibitem [{\citenamefont {Nayfeh}(1981)}]{nayfeh1981introduction}%
  \BibitemOpen
  \bibfield  {author} {\bibinfo {author} {\bibfnamefont {A.}~\bibnamefont
  {Nayfeh}},\ }\href {https://books.google.de/books?id=kzbvAAAAMAAJ} {\emph
  {\bibinfo {title} {Introduction to Perturbation Techniques}}},\ Wiley
  classics library\ (\bibinfo  {publisher} {Wiley},\ \bibinfo {year}
  {1981})\BibitemShut {NoStop}%
\bibitem [{\citenamefont {Goldobin}\ \emph {et~al.}(2017)\citenamefont
  {Goldobin}, \citenamefont {Pimenova}, \citenamefont {Rosenblum},\ and\
  \citenamefont {Pikovsky}}]{Goldobin2017}%
  \BibitemOpen
  \bibfield  {author} {\bibinfo {author} {\bibfnamefont {D.~S.}\ \bibnamefont
  {Goldobin}}, \bibinfo {author} {\bibfnamefont {A.~V.}\ \bibnamefont
  {Pimenova}}, \bibinfo {author} {\bibfnamefont {M.}~\bibnamefont {Rosenblum}},
  \ and\ \bibinfo {author} {\bibfnamefont {A.}~\bibnamefont {Pikovsky}},\
  }\href@noop {} {\bibfield  {journal} {\bibinfo  {journal} {The Eur. Phys. J.
  Spec. Topics}\ }\textbf {\bibinfo {volume} {226}},\ \bibinfo {pages} {1921}
  (\bibinfo {year} {2017})}\BibitemShut {NoStop}%
\bibitem [{\citenamefont {Rosenblum}\ \emph {et~al.}(2002)\citenamefont
  {Rosenblum}, \citenamefont {Pikovsky}, \citenamefont {Kurths}, \citenamefont
  {Osipov}, \citenamefont {Kiss},\ and\ \citenamefont
  {Hudson}}]{Rosenblum-Pikovsky-etal-02}%
  \BibitemOpen
  \bibfield  {author} {\bibinfo {author} {\bibfnamefont {M.}~\bibnamefont
  {Rosenblum}}, \bibinfo {author} {\bibfnamefont {A.}~\bibnamefont {Pikovsky}},
  \bibinfo {author} {\bibfnamefont {J.}~\bibnamefont {Kurths}}, \bibinfo
  {author} {\bibfnamefont {G.}~\bibnamefont {Osipov}}, \bibinfo {author}
  {\bibfnamefont {I.}~\bibnamefont {Kiss}}, \ and\ \bibinfo {author}
  {\bibfnamefont {J.}~\bibnamefont {Hudson}},\ }\href@noop {} {\bibfield
  {journal} {\bibinfo  {journal} {Phys. Rev. Lett.}\ }\textbf {\bibinfo
  {volume} {89}},\ \bibinfo {pages} {264102} (\bibinfo {year}
  {2002})}\BibitemShut {NoStop}%
\bibitem [{\citenamefont {Totz}\ \emph {et~al.}(2017)\citenamefont {Totz},
  \citenamefont {Rode}, \citenamefont {Tinsley}, \citenamefont {Showalter},\
  and\ \citenamefont {Engel}}]{Totz_etal-17}%
  \BibitemOpen
  \bibfield  {author} {\bibinfo {author} {\bibfnamefont {J.~F.}\ \bibnamefont
  {Totz}}, \bibinfo {author} {\bibfnamefont {J.}~\bibnamefont {Rode}}, \bibinfo
  {author} {\bibfnamefont {M.~R.}\ \bibnamefont {Tinsley}}, \bibinfo {author}
  {\bibfnamefont {K.}~\bibnamefont {Showalter}}, \ and\ \bibinfo {author}
  {\bibfnamefont {H.}~\bibnamefont {Engel}},\ }\href@noop {} {\bibfield
  {journal} {\bibinfo  {journal} {Nature Physics}\ }\textbf {\bibinfo {volume}
  {14}},\ \bibinfo {pages} {282} (\bibinfo {year} {2017})}\BibitemShut
  {NoStop}%
\bibitem [{\citenamefont {Gerson}\ \emph {et~al.}(2017)\citenamefont {Gerson},
  \citenamefont {Plagnol},\ and\ \citenamefont {Corr}}]{Gerson_etal-17}%
  \BibitemOpen
  \bibfield  {author} {\bibinfo {author} {\bibfnamefont {J.}~\bibnamefont
  {Gerson}}, \bibinfo {author} {\bibfnamefont {A.~C.}\ \bibnamefont {Plagnol}},
  \ and\ \bibinfo {author} {\bibfnamefont {P.~J.}\ \bibnamefont {Corr}},\
  }\href@noop {} {\bibfield  {journal} {\bibinfo  {journal} {Personality and
  Individual Differences}\ }\textbf {\bibinfo {volume} {117}},\ \bibinfo
  {pages} {81} (\bibinfo {year} {2017})}\BibitemShut {NoStop}%
\bibitem [{\citenamefont {Kuramoto}\ and\ \citenamefont
  {Battogtokh}(2002)}]{Kuramoto-Battogtokh-02}%
  \BibitemOpen
  \bibfield  {author} {\bibinfo {author} {\bibfnamefont {Y.}~\bibnamefont
  {Kuramoto}}\ and\ \bibinfo {author} {\bibfnamefont {D.}~\bibnamefont
  {Battogtokh}},\ }\href@noop {} {\bibfield  {journal} {\bibinfo  {journal}
  {Nonlinear Phenom. Complex Syst.}\ }\textbf {\bibinfo {volume} {5}},\
  \bibinfo {pages} {380} (\bibinfo {year} {2002})}\BibitemShut {NoStop}%
\bibitem [{\citenamefont {Zhang}\ \emph {et~al.}(2017)\citenamefont {Zhang},
  \citenamefont {Pikovsky},\ and\ \citenamefont {Liu}}]{Zhang_Pikovsky_Liu_17}%
  \BibitemOpen
  \bibfield  {author} {\bibinfo {author} {\bibfnamefont {X.}~\bibnamefont
  {Zhang}}, \bibinfo {author} {\bibfnamefont {A.}~\bibnamefont {Pikovsky}}, \
  and\ \bibinfo {author} {\bibfnamefont {Z.}~\bibnamefont {Liu}},\ }\href@noop
  {} {\bibfield  {journal} {\bibinfo  {journal} {Scientific Reports}\ }\textbf
  {\bibinfo {volume} {7}},\ \bibinfo {pages} {2104} (\bibinfo {year}
  {2017})}\BibitemShut {NoStop}%
\bibitem [{\citenamefont {Wolfrum}\ and\ \citenamefont
  {Omel'chenko}(2011)}]{Wolfrum-Omelchenko-11}%
  \BibitemOpen
  \bibfield  {author} {\bibinfo {author} {\bibfnamefont {M.}~\bibnamefont
  {Wolfrum}}\ and\ \bibinfo {author} {\bibfnamefont {O.~E.}\ \bibnamefont
  {Omel'chenko}},\ }\href@noop {} {\bibfield  {journal} {\bibinfo  {journal}
  {Phys. Rev. E}\ }\textbf {\bibinfo {volume} {84}},\ \bibinfo {pages} {015201}
  (\bibinfo {year} {2011})}\BibitemShut {NoStop}%
\bibitem [{\citenamefont {Kralemann}\ \emph {et~al.}(2008)\citenamefont
  {Kralemann}, \citenamefont {Cimponeriu}, \citenamefont {Rosenblum},
  \citenamefont {Pikovsky},\ and\ \citenamefont {Mrowka}}]{Kralemann-08}%
  \BibitemOpen
  \bibfield  {author} {\bibinfo {author} {\bibfnamefont {B.}~\bibnamefont
  {Kralemann}}, \bibinfo {author} {\bibfnamefont {L.}~\bibnamefont
  {Cimponeriu}}, \bibinfo {author} {\bibfnamefont {M.}~\bibnamefont
  {Rosenblum}}, \bibinfo {author} {\bibfnamefont {A.}~\bibnamefont {Pikovsky}},
  \ and\ \bibinfo {author} {\bibfnamefont {R.}~\bibnamefont {Mrowka}},\
  }\href@noop {} {\bibfield  {journal} {\bibinfo  {journal} {Phys. Rev. E}\
  }\textbf {\bibinfo {volume} {77}},\ \bibinfo {pages} {066205} (\bibinfo
  {year} {2008})}\BibitemShut {NoStop}%
\end{thebibliography}
\end{document}